\documentclass[twocolumn,showpacs,prb,preprintnumbers,amsmath,amssymb,groupedadress]{revtex4}
\usepackage{graphicx}
\usepackage{dcolumn}

\begin{document}
\title{Experimental time-resolved photoemission and \textit{ab initio} study \\of lifetimes of excited electrons in Mo and Rh}

\author{A. M\"{o}nnich}
\affiliation{Department of Physic, University of Kaiserslautern,
67663 Kaiserslautern, Germany}
\author{J. Lange}
\affiliation{Department of Physic, University of Kaiserslautern,
67663 Kaiserslautern, Germany}
\author{M. Bauer}
\affiliation{Department of Physic, University of Kaiserslautern,
67663 Kaiserslautern, Germany}
\author{M. Aeschlimann}
\affiliation{Department of Physic, University of Kaiserslautern,
67663 Kaiserslautern, Germany}
\author{I. A. Nechaev$^1$}
\altaffiliation[Also at: ]{Theoretical Physics Department, Kostroma
State University, 156961 Kostroma, Russia.}
\author{V. P. Zhukov$^{1}$}
\author{P. M. Echenique$^{1,2}$}
\author{E. V. Chulkov$^{1,2}$}
\affiliation{$^1$Donostia International Physics Center (DIPC), P. de
Manuel Lardizabal, 4,
20018, San Sebasti{\'a}n, Basque Country, Spain\\
$^2$Departamento de F{\'\i}sica de Materiales, Facultad de Ciencias
Qu{\'\i}micas, UPV/EHU and Centro Mixto CSIC-UPV/EHU, Apdo. 1072,
20080 San Sebasti\'an, Basque Country, Spain}

\date{\today}
\begin{abstract}
We have studied the relaxation dynamics of optically excited
electrons in molybdenum and rhodium by means of time resolved
two-photon photoemission spectroscopy (TR-2PPE) and \textit{ab
initio} electron self-energy calculations performed within the $GW$
and $GW+T$ approximations. Both theoretical approaches reproduce
qualitatively the experimentally observed trends and differences in
the lifetimes of excited electrons in molybdenum and rhodium. For
excitation energies exceeding the Fermi energy by more than 1 eV,
the $GW+T$ theory yields lifetimes in quantitative agreement with
the experimental results. As one of the relevant mechanisms causing
different excited state lifetime in Mo and Rh we identify the
occupation of the 4$d$ bands. An increasing occupation of the 4$d$
bands results in an efficient decrease of the lifetime even for
rather small excitation energies of a few 100 meV.

\end{abstract}
 \pacs{71.15.-m, 78.47.+p, 79.60.-i}
 \keywords{time-resolved photoemission, molybdenum, rhodium}
 \maketitle

\section{Introduction}
The dynamics of excited electrons play an important role for many physical and chemical phenomena. Detailed
knowledge of relaxation times of excited electrons is substantial for the interpretation of photoemission
spectra, surface chemistry data and low-energy electron diffraction data. For instance, the relaxation time
determines the inelastic electron mean free path, and hence the escape depth length of the applied surface
analytical technique.\cite{Penn} Particular in the optical excitation regime ($E < E_\text{Vac}$)
experimental data are very rare due to the limited experimental access to this energy regime. Therefore, the
knowledge of hot-electron lifetimes is still far from being complete. Among experimental methods developed
for such studies one of the most suitable is the technique of time-resolved two-photon photoemission
spectroscopy (TR-2PPE), which enables lifetime measurements directly in time domain at a temporal resolution
of a few femtoseconds\cite{Aesc96}. In the last decade several TR-2PPE experiments have been performed for
simple metals \cite{Bauer} and noble metals, \cite{Schm94,Hert96,Ogaw97,Knoes98,Caoo98} ferromagnetic 3$d$
metals (Fe, Co, and Ni), \cite{Aesc97,Knorr00} and high-$T_C$ superconductors \cite{Ness98}. Of the
paramagnetic transition metals, the electron lifetimes have been measured only for the 5$d$-metal
tantalum.\cite{Aesch04}. However, so far, no systematic experimental studies within the 4$d$ and
5$d$-transition metal series have been undertaken.

Theoretical calculations of excited electron lifetimes have been performed in the past within the $GW$
approximation (GWA) for electron self-energy for bulk noble \cite{Ech00,Zhuk01,Chu06} and 4$d$ transition
metals \cite{Chu06,Lads03,Lads04,Zhuk02a,Bace02} as well as for 5$d$ transition metal Ta.\cite{Aesch04} In
contrast to noble metals, which show qualitatively similar band structure and density of states
(DOS),\cite{Papaconst} electronic structure of 4$d$ metals varies strongly on moving from the beginning of
the 4$d$-series to the end. \cite{Papaconst} The calculations by Zhukov \textit{et al.} \cite{Zhuk02a} and
Bacelar \textit{et al.} \cite{Bace02} showed that the evaluated lifetimes also strongly vary along the
4$d$-series following trends in electronic structure. The extension, $GW+T$, of the $GW$ approximation by
inclusion of multiple
electron-hole scattering within a T-matrix approximation \cite{Spri98,Karl00,Zh05,%
NeChul05prb,NeChul05condmat} results in the decrease of the $GW$ lifetime value 
that brings theory and experiment to better agreement.
\cite{Aesch04,Zh05,NeChul05condmat}

In this paper we perform TR-2PPE measurements of excited electron lifetimes in Mo and Rh and compare these
results with the $GW$ and $GW+T$ calculated lifetimes. The choice of these metals is motivated by their very
distinct electronic structure. The Fermi level ($E_\text{F}$) of Mo, a typical $4d$ transition metal with bcc
crystalline structure, is located in a deep minimum of a wide and half-filled $d$-band (see Fig.
4).\cite{Papaconst} Rhodium in contrast, a 4$d$ metal with fcc crystalline structure, exhibits a very narrow
and almost-filled $d$-band \cite{Papaconst}, where the Fermi level is situated in the vicinity of a sharp
peak of density of states near the edge of the $d$-band (see Fig. 7). Hence, in comparison with molybdenum,
$d$ states of Rh are spatially more localized and, consequently, multiple scattering contributions to the
electron lifetime should be enhanced. Such considerable differences in the band structure between Mo and Rh
are expected to cause strong distinctions in the excited state lifetimes.

\section{\label{sec:experiment}Experiment}

With the TR-2PPE technique the decay dynamics of an electron excitation (intermediate state) located between
the Fermi edge $E_\text{F}$ and the vacuum level $E_\text{Vac}$ can be studied. The intermediate state is
populated by a first (pump) laser pulse. Due to electron-electron scattering processes the excited electrons
will then decay to lower unoccupied states. The rate of inelastic scattering of the electron defines the
inelastic lifetime of the intermediate state.
\begin{figure}[tbp]
\centering
\includegraphics[angle=90,scale=0.6]{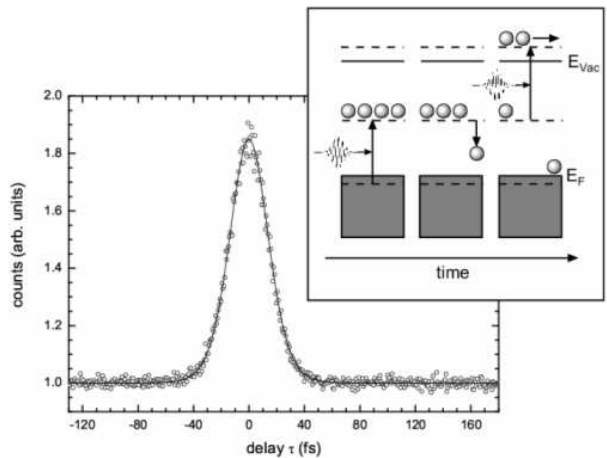}
 \caption{2PPE cross-correlation trace at an intermediate energy $E-E_\text{F}=1.3$\,eV; inset, principle
of a TR-2PPE pump and probe experiment.}
 \label{Mo_Auto}
\end{figure}%
After a controlled delay $\tau$ a second (probe) laser pulse transfers the remaining population to the
detected state above the vacuum level. The raw data (see Fig.~\ref{Mo_Auto}) of a TR-2PPE measurement
exhibits a 2-pulse correlation trace at a fixed kinetic energy, where the shape of the trace contains
information about the pulse width of the pump and the probe laser pulse as well as the population decay time
(and hence inelastic lifetime) of the probed intermediate state. In order to extract the intermediate state
lifetime from the crosscorrelation trace, a rate equation model of a three level system is solved within a
matlab-based fitting routine which uses a least-square-fit algorithm. Such a rate-equation model corresponds
to solutions of the Liouville-von Neumann equations within the density matrix in the limit of rapid
dephasing. It has been shown before to be a equivalent (and efficient) approach for the description of
electronic band excitations as investigated in the present studies \cite{Petek}.

The laser pulses for the TR-2PPE experiments were delivered by a mode-locked Ti:sapphire laser, pumped by
about 7\,W from a cw Nd:YVO$_3$ diode laser. The available energy range for 2PPE emission has been increased
by doubling the frequency of the linearly polarized pulses in a $0.2$ mm thick beta barium borate (BBO)
crystal.

To minimize coherent effects in the 2PPE signal the polarization of pump and probe pulse were set to $s$ and
$p$ polarized (cross polarized), respectively. The collinearly recombined beams were focused at $45^\circ$
incidence with respect to the surface normal of the sample. With these UV pulses ($h\nu$ = 3\,eV) of up to
$0.75$\,nJ, at an repetition rate of about 80MHz less than one electron per pulse was emitted from the sample
surface. In this means we avoid space-charge effects which can significantly distort the 2PPE signal.
Furthermore it guarantees that the excitation density of the intermediate excited state is kept at a low
level so that the experiment probes the relaxation of individual excited electronic states rather than the
collective behavior (thermalization dynamics) of a transiently heated nonequilibrium distribution. A
cylindrical sector energy analyzer (CSA)---positioned in the direction of the surface normal ($k_{\parallel}
= 0$)---was used to detect the electrons with an energy resolution of $100$\,meV.

For the non-time-resolved 2PPE measurements, we used a cw argon-ion-laser pumped mode-locked Ti:sapphire
laser delivering transform-limited pulses of about 120\,fs, $22.5$\,nJ/pulse at a repetition rate of 80\,MHz.
The advantage of this narrow bandwidth laser system is its tunability in the laser wavelength within the
amplification bandwidth of about 100 nm which is required to distinguish between spectral features in the
2PPE spectra, arising from the occupied and unoccupied density of states of the electronic structure (see
below).\cite{Fauster} The wavelength of the laser pulses can be varied from 740 to 840\,nm which corresponds
to a tuning range of $3.35$ to $2.95$\,eV in the photon energy of the frequency-doubled laser pulses.

\begin{figure}[tbp]
\centering
\includegraphics[angle=0,width=0.8\linewidth]{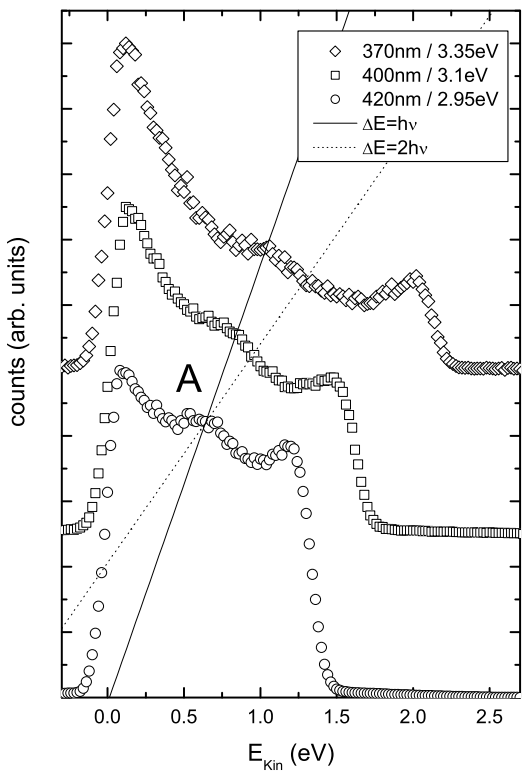}
 \caption{2PPE spectra of molybdenum recorded at varying photon energy; solid line,
peak shift as expected for an unoccupied state; dotted line, peak shift as expected of an occupied state.}
 \label{Mo_peak}
\end{figure}%
Metallic Mo and Rh sheets of approximately $0.1$-mm thickness were used for our experiments. We chose
polycrystalline samples to guarantee an averaging over all possible directions in the $k$ space of the
Brillouin zone and to avoid any influence of surface resonances on the measured lifetime which can exceed the
bulk lifetime values by a factor of 10 and more.\cite{Fauster} The purity of the sample was specified to be
higher than $99.9$\%. The metallic sheets were cleaned by several sputter and heating cycles. After the
cleaning we found no indication of significant impurities, such as C or O at the surface. Next to the sheet
samples further TR-2PPE experiments were performed at ultrathin polycrystalline films of molybdenum of
varying thickness between 20 nm and 100 nm. The films were prepared in-situ  by means of electron-beam
evaporation on a dielectric MICA substrate from a molybdenum rod (purity, 99.9+\%) at a pressure of
$<10^{-9}$ mbar. The film thickness was controlled using a calibrated flux monitor.

The cleaned metal surface was dosed with a small amount of Cs ($<0.1$\,ML) to lower the surface work
function. This procedure enables to increase the accessible energy range in the 2PPE experiment by about $1$
eV so that excitation levels also very close to the Fermi energy can be probed. Cs is evaporated from a
thoroughly outgassed commercial getter source. The effect on the lifetime measurements by dosing a metal
surface with a small amount of Cs has been thoroughly examined in the past \cite{Baue99}. Using
cross-polarized pulses no differences in the lifetime of volume states could be observed between a clean and
a cesiated metal surface by means of TR-2PPE.

Static 2PPE spectra of an uncesiated molybdenum surface recorded at different photon energies are displayed
in Fig.\ \ref{Mo_peak}. In accordance with conventional photoemission experiments, the position of the
low-energy cutoff of the spectra is governed by the work function of the surface. The sharp high-energy
cutoff on the other hand maps the Fermi edge of the metallic sample. 2PPE spectra from polycrystalline
samples are typically characterized by a rather structureless spectrum due to the restricted spectral width
accessible with the used excitation wavelength in the optical regime and due to a blurring of spectral
features by the involvement of an intermediate state in the photoemission process. The appearance of a
distinct peak (indicated by A) at $E_\text{Kin}\approx0.6$\,eV (corresponding to an intermediate state energy
of $E-E_\text{F}= 2.2$ eV) in the present example is insofar somewhat unusual. Rhodium, for instance,
exhibits a completely structureless 2PPE spectrum. According to the excitation process, the shift in the peak
position by $\Delta E=h\nu$ as the excitation wavelength is tuned, indicates that the peak is due to an
feature in the unoccupied DOS of Mo. In contrast, an occupied state feature would be characterized by an
energy shift proportional to $\Delta E=2h\nu$ (see dotted line in Fig.\ \ref{Mo_peak} and the shift of the
Fermi edge). Our measurements are performed on a polycrystalline sample and, furthermore, we do not observe
any dependence in the appearance of the peak as a function of the excitation light polarization. We,
therefore assign, this peak to an increase in the volume density of states in the corresponding intermediate
energy region rather than to a surface state. Our band structure calculations presented in Sec.\
\ref{subsec:results_Mo}, show indeed a drastic increase in the DOS at about $2.25$\,eV above the Fermi level
(see Fig. 4).

\section{\label{sec:theory}Theory}
Within the framework of many-body perturbation theory,\cite{Inkson,FettWal} the inverse lifetime (decay rate)
of an excited electron (hole) in a single-particle Bloch state $\psi_{\mathbf{q}n}$ with eigenvalue
$\epsilon_{\mathbf{q}n}$ is given by\cite{Zhuk02a}
\begin{equation}\label{lifetime}
\tau^{-1}_{\mathbf{q}n}=2|\mathrm{Im}\Delta\epsilon_{\mathbf{q}n}|,
\end{equation}
where $\Delta\epsilon_{\mathbf{q}n}$ is the many-body self-energy corrections to $\epsilon_{\mathbf{q}n}$. In
the Dyson scheme, these corrections can be approximated as\cite{Zhuk01,Zhuk02a}
\begin{equation}\label{SE_corrections}
\Delta\epsilon_{\mathbf{q}n}=Z_{\mathbf{q}n}\Delta\Sigma_{\mathbf{q}n}(\epsilon_{\mathbf{q}n}),
\end{equation}
where $Z_{\mathbf{q}n}$ is the renormalization factor related to the partial derivative of
$\Delta\Sigma_{\mathbf{q}n}(\omega)$ with respect to energy at $\omega=\epsilon_{\mathbf{q}n}$. Here
$\Delta\Sigma_{\mathbf{q}n}(\omega)$ is the matrix elements
$\langle\psi_{\mathbf{q}n}|\Delta\Sigma(\omega)|\psi_{\mathbf{q}n}\rangle$ of the difference
$\Delta\Sigma(\omega)=\Sigma(\omega)-v^{xc}_{LDA}$ between the self-energy $\Sigma(\omega)$ and the exchange
and correlation potential obtained in the local-density approximation (LDA). Thus, the main problem in the
lifetime evaluation is the calculation of the matrix elements of the self-energy
$\Sigma_{\mathbf{q}n}(\omega)$.

\begin{figure}[tbp]
\centering
 \includegraphics[angle=-90,scale=1.0]{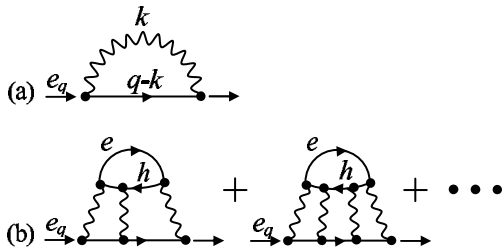}
\caption{A schematic representation of the $GW$ self-energy (a) and the $T$-matrix contribution (b). The
wiggly lines signify the RPA dynamically screened Coulomb interaction $W^0$. The solid lines with arrows
represent the Green function $G^0$.}\label{diagrams}
\end{figure}

A widely used approximation to calculate the self-energy is the GWA.\cite{GW_review} This approximation is
formally the first-order term in the Hedin expansion\cite{Hedin} of the self-energy in power of the screened
Coulomb interaction. In the simplest version of the GWA we consider here, the self-energy is defined by the
product of the \textit{noninteracting} Green function $G^0$ and the screened interaction $W^0$ obtained
within the \textit{random phase approximation} (RPA).\cite{Inkson,FettWal,FerdiAnis} Drawing attention only
on momentum and energy dependence with the use of a shorthand notation $q=(\mathbf{q},\omega)$ in order to
simplify the presentation, we can write the $GW$ self-energy $\Sigma_{\mathbf{q}n}^{GW}(\omega)$
as\cite{Hedin}
\begin{equation}\label{GW_Sigma}
\Sigma^{GW}(q)=\frac{i}{(2\pi)^4}\int\,dkG^0(q-k)W^0(k).
\end{equation}
Both $G^0$ and $W^0$ are built on the $\{\psi_{\mathbf{k}n},\epsilon_{\mathbf{k}n}\}$ set. Physically
Eq.~(\ref{GW_Sigma}) means that an electron, $e_q$, propagating in a system with the momentum $q$, creates a
disturbance carrying momentum $k$ [see Fig.~\ref{diagrams}(a)]. This fact is reflected in the formula by the
interaction $W^0(k)$. Further, as depicted by $G^0(q-k)$ in Eq.~(\ref{GW_Sigma}), the electron propagates
with the momentum $(q-k)$ right up to absorbtion of the disturbance.

Using the spectral function representation of $G^0$ and $W^0$ in Eq.~(\ref{GW_Sigma}), we can explicitly
write down the imaginary part of the $GW$ self-energy for, e.g., excited electrons ($\omega>\epsilon_F$) as
\begin{eqnarray}\label{Sigma_e}
\mathrm{Im}\Sigma_{\mathbf{q}n}^{GW}(\omega)&=&-\sum_{\mathbf{k}}\sum_{n'>\epsilon_F}\sum_{ij}
\langle\psi_{\mathbf{q}n}\psi_{\mathbf{k}-\mathbf{q}n'}|B_{\mathbf{k}i}\rangle \nonumber\\
&\times&\mathrm{Im}W^{0}_{ij}(\mathbf{k},\omega-\epsilon_{\mathbf{k}-\mathbf{q}n'}) \nonumber \\
&\times&\langle B_{\mathbf{k}j}|\psi_{\mathbf{k}-\mathbf{q}n'}\psi_{\mathbf{q}n}\rangle
\Theta(\omega-\epsilon_{\mathbf{k}-\mathbf{q}n'}).
\end{eqnarray}
Here $\epsilon_F$ is the Fermi energy, $\Theta(x)$ is the step function, $\{B_{\mathbf{k}i}(\mathbf{r})\}$ is
a set of appropriate Bloch basis functions, and $W^0_{ij}$ are matrix elements of the RPA screened
interaction in this basis. In the present work, one-electron wave functions $\psi_{\mathbf{q}n}$, energies
$\epsilon_{\mathbf{q}n}$, and $B_{\mathbf{k}i}$ are calculated by using the linear muffin-tin orbital (LMTO)
method.\cite{Andersen} Having obtained the imaginary part of the self-energy, its energy-dependent real part
can be found from the Hilbert transform.  For details, we refer the reader to Refs.~\onlinecite{FerdiAnis}
and \onlinecite{Zhuk01}.

By making use of the $T$ matrix that describes multiple scattering between an electron and a hole, we can go
beyond the GWA in the calculation of the self-energy. In this case, we \textit{additionally} take into
account third- and higher-order terms in the Hedin expansion of the self-energy which constitute an infinite
class of electron-hole ladder diagrams.\cite{Inkson,NeChul05prb} This allows us to consider the following
scattering processes [see Fig.~\ref{diagrams}(b)]. The $e_q$ electron creates a disturbance which in its turn
results in the creation of an electron($e$)-hole($h$) pair. After that, the propagating electron repeatedly
interacts (one or more times) with the hole during the $e$-$h$-pair transport until this pair annihilation.
The latter causes the disturbance which further is adsorbed by $e_q$ at the end of the considered process.

Without going into details which one can find in Refs.~\onlinecite{NeChul05prb} and
\onlinecite{NeChul05condmat}, we write the $T$-matrix contribution as an \textit{additional} term to the $GW$
self-energy of Eq.~(\ref{GW_Sigma}):
\begin{equation}\label{T_matrix_term}
\Sigma^{T}(q)=\frac{i}{(2\pi)^4}\int\,dkG^0(q-k){\cal T}(k),
\end{equation}
where the $T$ matrix is given by
\begin{equation}\label{T_matrix_3O}
{\cal T}(q)=\widetilde{W}(q)
P^0(q)\widetilde{W}(q)\left(\frac{1}{1+\frac{1}{2}P^0(q)\widetilde{W}(q)}-1\right).
\end{equation}
The local electron-hole interaction $\widetilde{W}(q)$ is simply related with the first-order exchange
correction to $P^0(q)$ accounting for a single electron-hole scattering event. Note that
Eq.~(\ref{T_matrix_3O}) should be considered as a relation between the matrices ${\cal
T}_{ij}(\mathbf{q},\omega)$, $\widetilde{W}_{ij}(\mathbf{q},\omega)$, and $P^0_{ij}(\mathbf{q},\omega)$. It
is important that the $T$-matrix contribution of Eq.~(\ref{T_matrix_term}) has a $GW$-like formula. This
means that in order to obtain the imaginary part of $\Sigma^{T}_{\mathbf{q}n}(\omega)$, we can use Eq.
(\ref{Sigma_e}) with ${\cal T}_{ij}$ instead of $W^{0}_{ij}$.

The resulting self-energy in such a $GW+T$ approach is given by the sum
$\Sigma^{GW+T}_{\mathbf{q}n}(\omega)=\Sigma^{GW}_{\mathbf{q}n}(\omega)+\Sigma^{T}_{\mathbf{q}n}(\omega)$. In
calculations presented below, following Refs.~\onlinecite{Karl00,Zhukov}, and \onlinecite{Aesch04}, we model
the interaction $\widetilde{W}$ as the Fourier transform of the RPA screened interaction $W^0$ at zero energy
but with the $2/3$ prefactor obtained in Ref.~\onlinecite{NeChul05condmat}, i.e.,
$\widetilde{W}_{ij}(\mathbf{k},\omega)=\frac{2}{3}W^{0}_{ij}(\mathbf{k},0)$. Having obtained the self-energy,
one can evaluate the lifetime with the help of Eqs. (\ref{lifetime}) and (\ref{SE_corrections}). In order to
compare theoretical results with the experimental data (polycrystalline sample), we average the calculated
lifetime over momentum for a given excitation energy.
\section{\label{sec:results}Results and discussion}
\subsection{\label{subsec:results_Mo}Molybdenum}
\begin{figure}[tbp]
\centering
 \includegraphics[angle=-90,scale=0.33]{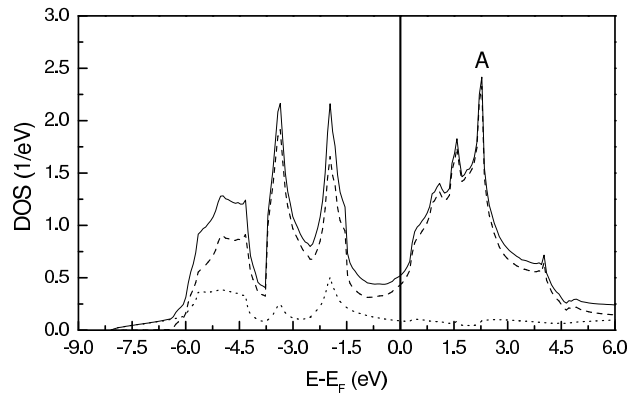}
\caption{Total and partial DOS in Mo. The total DOS is shown by solid line, the 4d DOS by dashed line, and
the 5s+5p DOS by dotted line.} \label{bnd_dos_mo}
\end{figure}%
\begin{figure}[tbp]
\centering
 \includegraphics[angle=-90,scale=0.33]{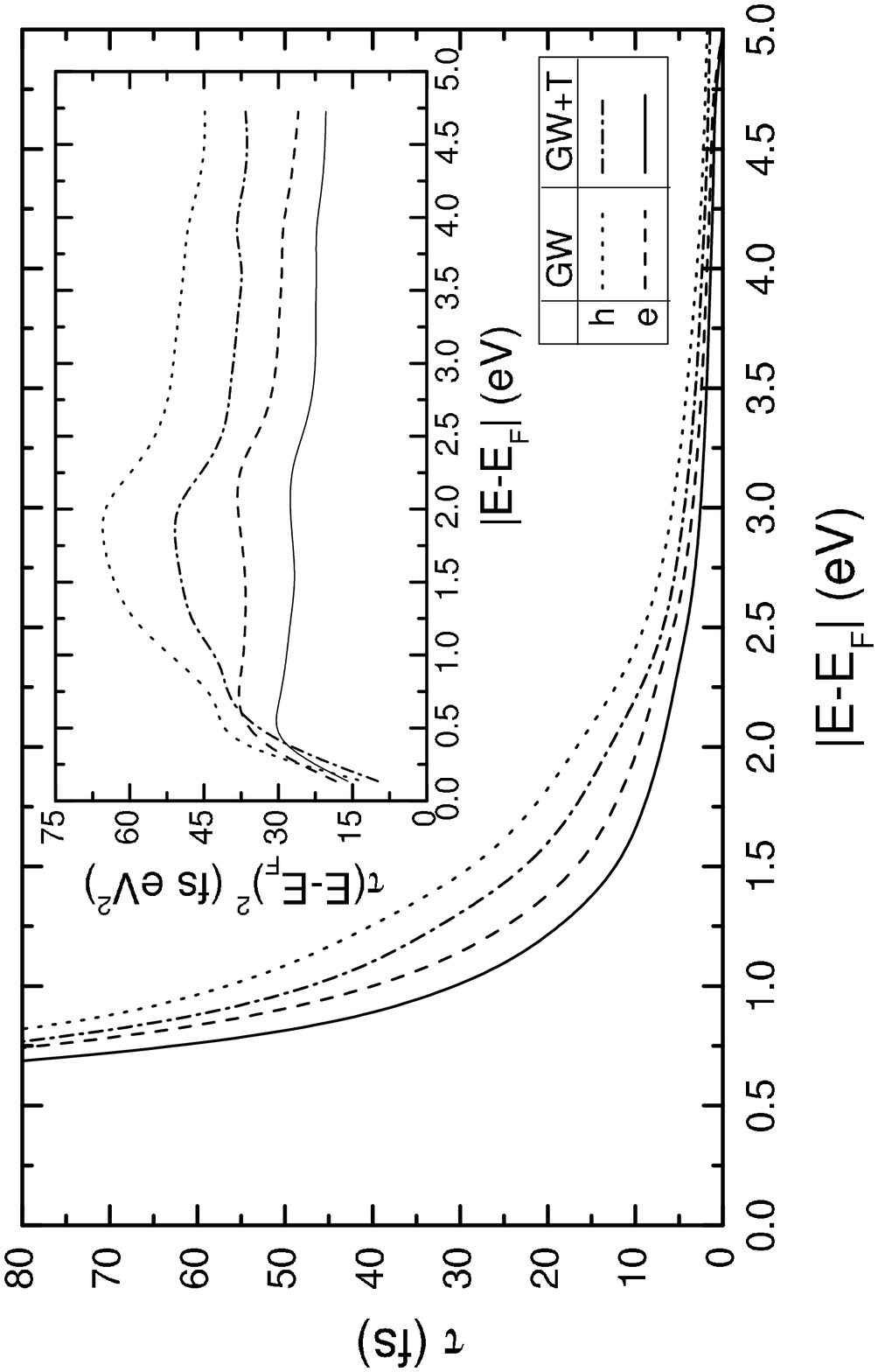}
\caption{Momentum-averaged lifetimes $\tau$ of excited electrons (e) and holes (h) in Mo calculated within
$GW$ and $GW+T$ approximations. Inset, scaled momentum-averaged electron and hole lifetimes,
$\tau(E-E_\text{F})^2$, in Mo.} \label{tau_e_h_full}
\end{figure}%
\begin{figure}[tbp]
\centering
  \includegraphics[width=1\linewidth]{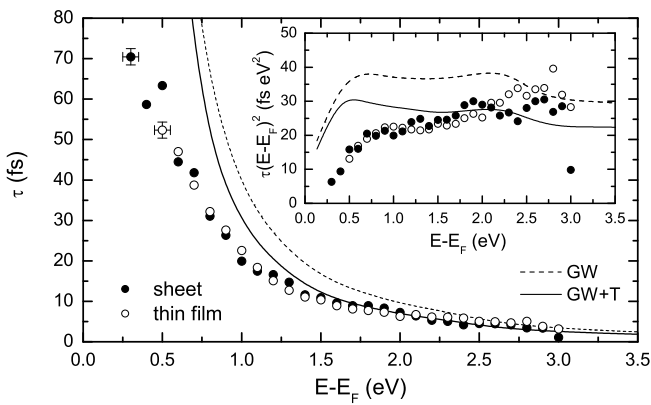}
\caption{Calculated and experimental momentum-averaged electron lifetimes $\tau$ in Mo. Open circles are
lifetimes obtained from a polycrystalline Mo thin film (thickness $\approx 20$ nm), filled circles are
lifetimes measured on a polycrystalline Mo sheet. Inset, scaled momentum-averaged electron and hole
lifetimes, $\tau(E-E_\text{F})^2$, in Mo.} \label{tau_part_e}
\end{figure}%

Figure~\ref{bnd_dos_mo} shows the band structure of Mo and corresponding total and partial DOS. As mentioned
before, the DOS has a wide ($\approx11$ eV) $d$-band with a deep minimum at $E_\text{F}$. Until $\sim5$~eV
above $E_\text{F}$ the total DOS is almost completely determined by the unoccupied $d$-states and has a sharp
peak at 2.25 eV. This peak can be identified with the experimentally observed unoccupied peak A discussed in
Sec.~\ref{sec:experiment}. Below the Fermi level, the $d$ states also dominate in the formation of the total
DOS. At the same time, as it is evident from Fig.~\ref{bnd_dos_mo}, the contribution of $sp$ states is
essentially larger for these energies than for energies above $E_\text{F}$. The total DOS shows two
pronounced peaks (at -1.96 eV and at -3.4 eV) separated by a minimum at -2.52 eV.

Figure~\ref{tau_e_h_full} shows the calculated momentum-averaged lifetimes of electrons and holes. As follows
from the figure, the electron lifetime is shorter than the hole one. It is caused by both the considerable
$sp$ contribution below the Fermi level and the total DOS shape which differs from that above $E_\text{F}$
(the available phase space for hole inelastic scattering is smaller). Nevertheless, as can be seen from the
inset in Fig.~\ref{tau_e_h_full}, due to the fact that the total DOS has a relatively symmetric form with
respect to $E_\text{F}$ the electron and hole $GW$ lifetimes demonstrate a quite similar behavior. Also, both
lifetimes have a small knoll in the vicinity of the excitation energy of $\sim$2 eV arising from the key
features of the electron structure of Mo. As evident from the figure, the inclusion of the $T$ matrix leads
to a shortening of the lifetimes. An analysis of the ratio
$\gamma=\mathrm{Im}\Sigma^{GW+T}/\mathrm{Im}\Sigma^{GW}$ has shown that in the vicinity of $E_\text{F}$
($\pm0.5$ eV) the $T$-matrix contribution is essentially less than the $GW$ term. However,
$\mathrm{Im}\Sigma^{T}$ increases rather fast with increasing excitation energy and amounts on average to
$\sim35\%$ with respect to $\mathrm{Im}\Sigma^{GW}$ at $|E-E_\text{F}|\approx2$ eV.

Addressing the role of $d$ states in the formation of the energy dependence of the lifetime, it is worth
emphasizing that taking into account only $d$ states in the evaluation of the $e-h$ propagator $P^0$ and the
Green function in the GWA, one obtains similar results for both the electron and hole
lifetimes.\cite{Unpublished_calcs} This means that the $d$-band plays a dominant role and only 4$d$ states
can be treated to quickly assess the imaginary part of the self-energy and the lifetime.

In Fig.~\ref{tau_part_e}, the calculated momentum-averaged electron lifetimes are presented in comparison
with the experimentally determined relaxation times of photoexcited electrons extracted from the
cross-correlation traces. The highest excited state probed in the experiments was restricted by the photon
energy to $3.0$\,eV. The transport of excited electrons out of the surface region into the bulk can act as an
additional decay channel which reduces the measured lifetime of the excited states.\cite{Aesc00} In order to
study the influence of transport effects on the experimental results, we performed further experiments on the
carefully prepared Mo films suported by a dielectric MICA substrate. Different film thickness between 20\,nm
and 100\,nm were studied. As evident from Fig.~\ref{tau_part_e}, both, sheet data (solid circles) and thin
film data (open circle), give identical results within the scatter of the data. The lack of any systematic
deviation shows that the contribution of transport to the lifetime measured in the TR-2PPE experiment is
negligible.\cite{Aesc00,Aesch04}

Regarding the comparison between experimental and theoretical data, inspecting Fig.~\ref{tau_part_e} one can
see that the $T$-matrix inclusion  leads to better agreement between the theoretical and experimental results
(especially for $E-E_\text{F}\gtrsim1.0$ eV). This becomes particularly evident in the inset of
Fig.~\ref{tau_part_e} which shows the deviation of experimental and calculated lifetime data from the
Fermi-liquid model by Quinn and Ferrell\cite{QF}. Here, the center of gravity of the experimental scaled
lifetime ($\approx25$ fs eV$^2$) is close to the $GW$+$T$ curve for energies higher than $\sim0.7$\,eV. Below
this value, towards $E_\text{F}$, the experimental lifetime goes down faster than the $GW+T$ one. However,
the experimentally observed less rapid increase in the lifetime towards $E_\text{F}$ in comparison to the
Fermi-Liquid model in this energy region is qualitatively well reproduced by the theoretical results, which
follow the slope of the data reasonably well. Some discrepancy for energies smaller than $1.0$ eV can be
caused by both the approximations made for the self-energy (e.g., electron-phonon coupling can be important
here) and secondary processes which effect the experimental observed relaxation times.\cite{Aesc00} Note that
this discrepancy can be eliminated by adding to $\mathrm{Im}\Sigma^{GW+T}$ a weak energy dependent term of
$\sim5$ meV.

\subsection{\label{subsec:results_Rh}Rhodium}
\begin{figure}[tbp]
\centering
 \includegraphics[angle=-90,scale=0.33]{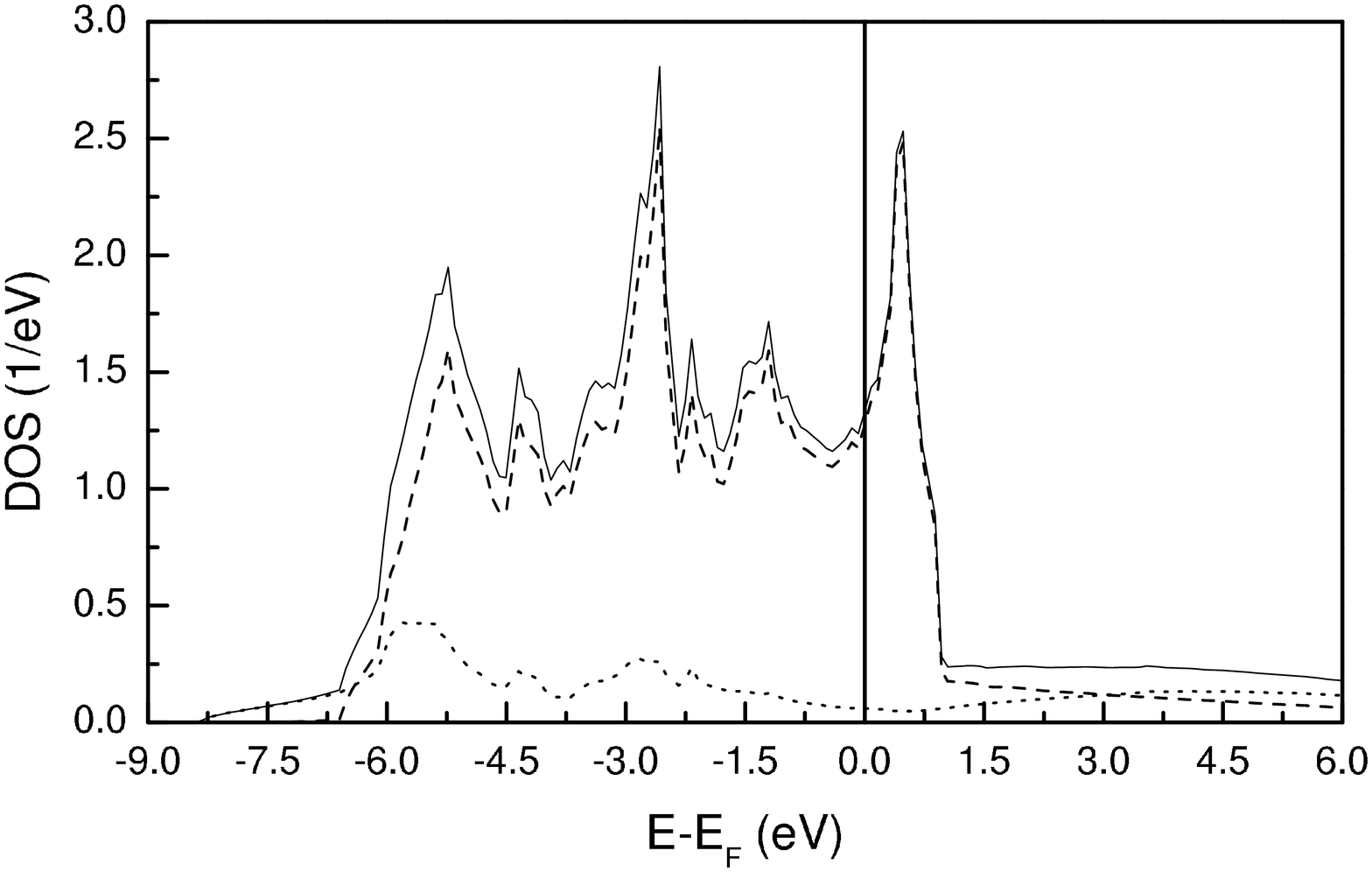}
\caption{Total and partial DOS in Rh. Total DOS is shown by solid line, the 4$d$ DOS by dashed line, and the
5$s$+5$p$ DOS by dotted line.}\label{bnd_dos_Rh}
\end{figure}%
\begin{figure}[tbp]
\centering
 \includegraphics[angle=-90,scale=0.33]{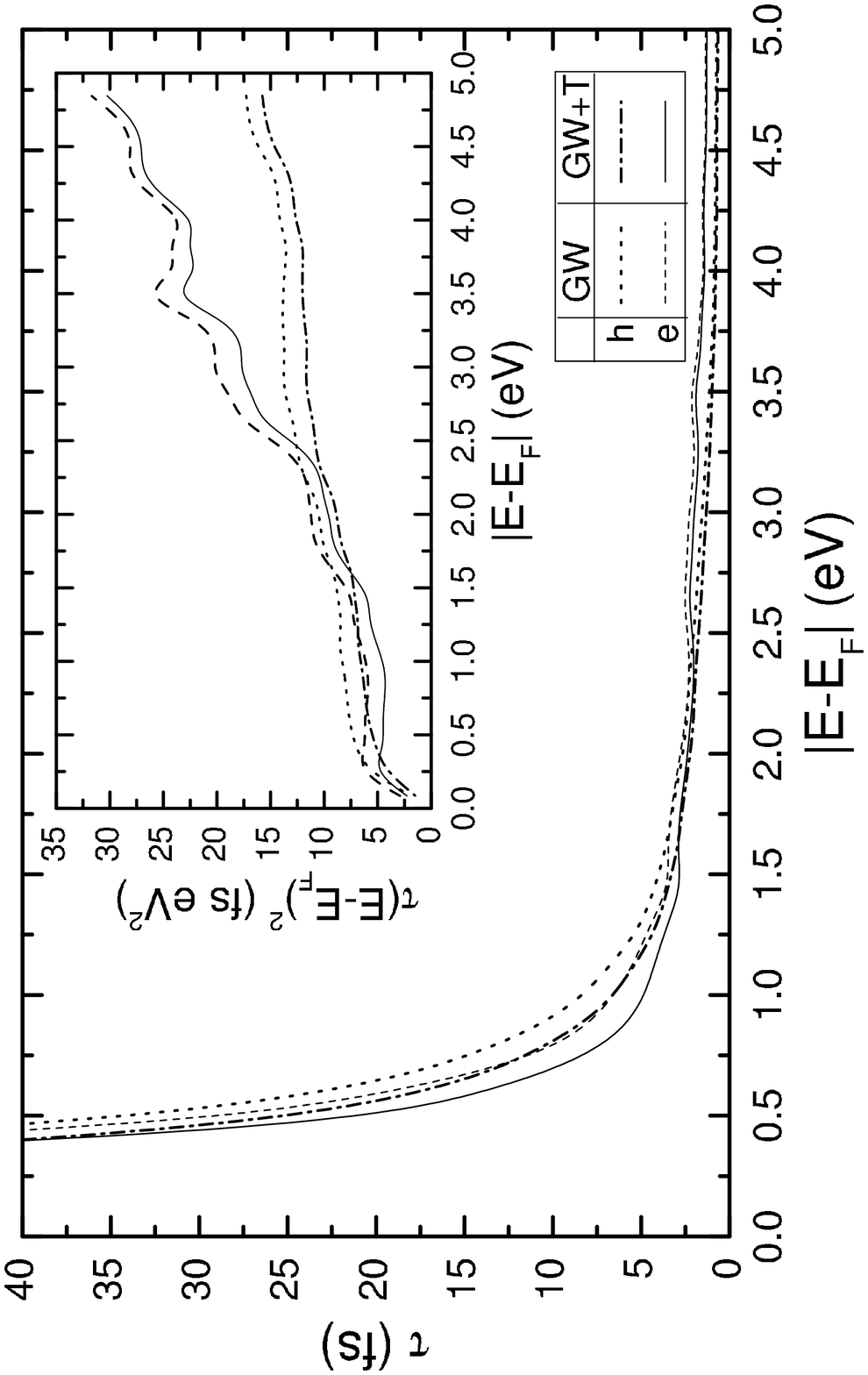}
\caption{Momentum-averaged lifetimes $\tau$ of excited electrons (e) and holes (h) in Rh calculated within
$GW$ and $GW+T$ approximations. Inset, scaled momentum-averaged electron and hole lifetimes,
$\tau(E-E_\text{F})^2$, in Rh.}\label{tau_teor_Rh}
\end{figure}%
\begin{figure}[tbp]
\centering
 \includegraphics[width=1\linewidth]{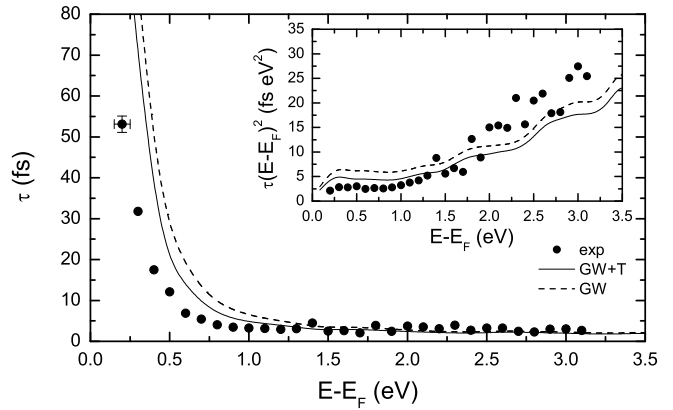}
\caption{Calculated and experimental momentum-averaged electron lifetimes $\tau$ in polycrystalline Rh.
Inset, scaled momentum-averaged electron lifetime in Rh.}\label{tau_part_e_Rh}
\end{figure}%

The calculated band structure and DOS of Rh are shown in Fig.~\ref{bnd_dos_Rh}. Compared to Mo, the Rh
$d$-band is more narrow ($\sim7$ eV). Almost all the $d$ states are occupied except for a small fraction
which constitutes the sharp peak at $0.46$ eV and pronounced edge of the $d$-band at $0.9$ eV above
$E_\text{F}$. At higher energies unoccupied bands show comparable contributions of $sp$ and $d$ states. Below
$E_\text{F}$, the $sp$-states contribution to the bands is negligible in comparison with the $d$ states.
Another peak comparable with the previous one is situated at $-2.58$ eV.

In Fig.~\ref{tau_teor_Rh}, the calculated momentum-averaged lifetime of electrons and holes in Rh is shown.
On first glance, it may seem that the lifetimes of electrons and holes demonstrate a similar behavior and
differ to only a small extent. However, inspecting the scaled averaged lifetimes (see inset in
Fig.~\ref{tau_teor_Rh}), one can see that starting from an energy corresponding to the $d$-band edge, the
energy dependence of $\tau(E-E_\text{F})^2$ for electrons can be fitted by a straight line with the slope of
$\approx6.8$ fs eV against $\approx2.2$ fs eV for the hole scaled lifetime. Such a distinction is caused by
the total DOS shape, the main part of which lies below $E_\text{F}$. Note that at energies below the $d$-band
edge both the considered curves exhibit the scaled lifetime energy dependence similar to that in the case of
Mo, but with essentially smaller values of $\tau(E-E_\text{F})^2$ due to the higher DOS.\cite{Zarate}

As in Mo, the inclusion of the $T$ matrix leads to a shortening of the lifetime. However, in this case, in
the vicinity of $E_\text{F}$ ($\pm0.4$ eV) the $T$ matrix yields the largest contribution which amounts on
average to $\sim40\%$ in relation to $\mathrm{Im}\Sigma^{GW}$. As the excitation energy increases, the
contribution falls down. At $|E-E_\text{F}|\approx3.0$ eV, the ratio $\gamma$ has knolls running up to $1.13$
for electrons and $1.17$ for holes. Calculations performed with taking into account only $d$ states have
corroborated\cite{Unpublished_calcs} the dominant role of the $d$-band in formation of the excited electrons
dynamics in transition metals.

In Fig.~\ref{tau_part_e_Rh}, we show the experimental relaxation time in comparison with the calculated
momentum-averaged electron lifetimes. Again, as it has been done for Mo, the highest excited state probed in
the experiments was restricted by the photon energy to $3.0$\,eV. The lifetime measurements for
polycrystalline thin films were not carried out in this case. Taking into consideration that the relaxation
times of excited electrons in Rh are rather short over a large energy region, we have supposed that transport
effects are negligible. Actually, the considered quantity remains below $7$ fs down to an exciting energy
$E-E_\text{F}\approx0.7$ eV (see experimental points in Fig. \ref{tau_part_e_Rh}). Towards the Fermi energy a
steep increase in the relaxation time is obtained. This is the lowest onset energy in a lifetime increase
ever reported in TR-2PPE measurements of bulk states.

Comparing the experimental and theoretical data shown in Fig.~\ref{tau_part_e_Rh}, one can distinguish the
following main aspects. The inclusion of the $T$ matrix ensures better agreement between the theoretical and
experimental results for $E-E_F\lesssim1.7$ eV (see the inset), whereas for higher energies the experimental
values are even above the $GW$ curve. As in Mo, some discrepancy in the low energy region can be eliminated
by adding to $\mathrm{Im}\Sigma^{GW+T}$ a weak energy dependent term, but the latter is equal to $\sim10$ meV
in the case of Rh.

\subsection{\label{subsec:comparison}Comparison  of lifetimes in  Mo and Rh  }

\begin{figure}[tbp]
\centering
 \includegraphics[angle=0,width=0.8\linewidth]{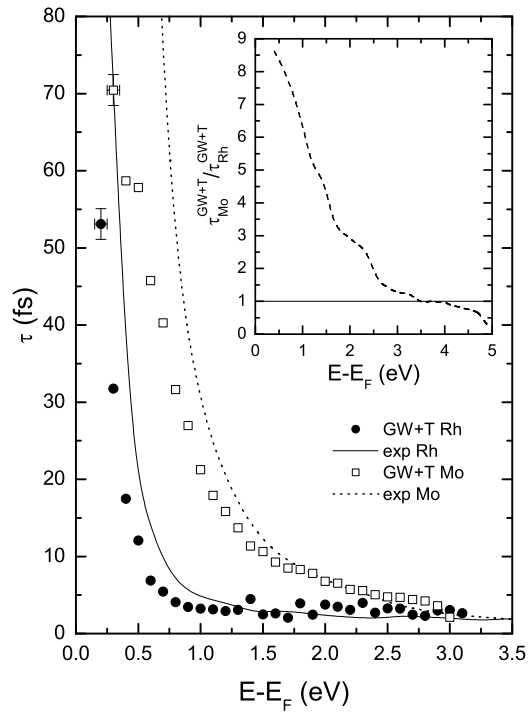}
\caption{Calculated averaged electron lifetimes $\tau$ and experimental relaxation time in polycrystalline Rh
and Mo. Inset, ratio of calculated electron lifetimes in Mo and Rh.}\label{Rh_Mo_theo_exp}
\end{figure}

In Fig.~\ref{Rh_Mo_theo_exp} we compare the time-resolved 2PPE data as well as the theoretical results for
the $4d$ transition metals rhodium and molybdenum. With a glance at the figure, practically over all the
considered energy region the electron lifetimes in Rh are shorter than those in Mo, especially below 2.0 eV.
Presented as the ratio $\tau^{GW+T}_{\mathrm{Mo}}/\tau^{GW+T}_{\mathrm{Rh}}$ (see inset in
Fig.~\ref{Rh_Mo_theo_exp}), the calculated $GW+T$ lifetimes, which are consistent with the experimental data,
clearly reflect this fact. Moreover, one can see that at energies above 1.0 eV, where agreement between the
experimental and theoretical results is rather good, this ratio falls down from 6 to unity (at $\sim3.5$ eV).
Further, the ratio becomes smaller than 1. By this comparison we demonstrate that upon moving across the $4d$
series from Mo to Rh the changes in filling, width, and shape of the $d$-band have a strong effect on the
electron lifetime.

In principle, such a shortening of the lifetime is an expected trend which can be qualitatively understood in
a scattering theory approach (STA).\cite{Zarate,Ech00,Zhuk02a} Within this approach, the lifetime is
considered in terms of energy-dependent matrix element $|M(\omega)|^2$ and the total DOS convolution. The
former contains effects of screening, and the latter reflects peculiarities of one-electron energy
dispersion. As was shown in Ref.~\onlinecite{Zhuk02a}, for Mo and Rh a good accordance between the STA and
GWA calculations can be achieved with the energy-independent matrix element $|M|^2$. Furthermore, for excited
electrons the matrix elements for Mo and Rh are similar. Therefore, an analysis of differences between the
lifetimes in these transition metals can be reduced to an examination of features of the DOS. In addition,
due to the dominant role of $d$ states, only the partial $d$-DOS is of interest. For example, taking into
account the density of $d$ states averaged over the $d$-band width, as we move from Mo to Rh, one can obtain
the electron lifetime decreasing by a factor of 3. Further, considering  molybdenum as a representative of
bcc metals from the center of the $d$-transition series and rhodium as a representative of fcc metals from
the end of these series, one can expect approximately the same shortening of the electron lifetime upon
moving, e.g., from Nb to Pd (the $4d$ series) and from Ta or W to Pt (the $5d$
series).\cite{Unpublished_calcs}

\section{\label{sec:conclusions}Conclusions}
In conclusion, we have analyzed the strong influence of the transition metal 4$d$-band structure on the
relaxation dynamics of optically excited electrons. Comparing the obtained theoretical results on Mo and Rh
with the experimental data, we have shown that the $T$-matrix contribution to the electron self-energy
improves the GWA values. Thus, we have found that taking into account multiple scattering between an excited
electron and a hole is important to achieve reasonable agreement with the TR-2PPE spectroscopy measurements.
Next to the good agreement between experimental and theoretical lifetime data we were also able to assign a
feature in the static molybdenum 2PPE spectrum to an enhanced electronic density of the excited state
distribution in the calculated bulk band structure. Overall we believe that the present study is a good
example of how the combination of sophisticated state-of-the-art experimental and theoretical approaches
enable a microscopic understanding of the ultrafast femtosecond dynamics of electronic excitation in complex
condensed matter systems.

\section*{Acknowledgments}
This work was partially supported by Gobierno Vasco, UPV/EHU, MCyT (Grant No. FIS 2004-06490-C03-01), by the
European Community 6th framework Network of Excellence NANOQUANTA (Grant No. NMP4-CT-2004-500198) and by the
priority program 1133 of the German Science Foundation (DFG).


\end{document}